# Lagerkvist versus Crick

Denis Semyonov.

Novosibirsk State University, 630090, Novosibirsk, Russia

Tel.:

*E-mail address*: dasem@mail.ru

# ABSTRACT

Present day data allow significant reconsideration of ideas on mechanisms underlying the degeneracy in the genetic code. Here a hypothesis is presented which links the degeneracy to possible conformational alterations in the codon-anticodon duplex. This enables explanation of Rumer symmetry in the table of the genetic code, coding of methionine and tryptophane without degeneracy and even predict significant difference between tautomer features of thymine and uracyl. The suggested hypothesis has something in common with Lagerkvist's idea that only two nucleotides in codon are coding.

# 1. Introduction

One of the keystone processes of the life is protein synthesis on ribosomes, which is translation of the genetic information from four-letter nucleotide language of nucleic acids to twenty-letter amino acid language of polypeptides (proteins). The central conception summarizing ideas on the rules of translation is the genetic code representing a correspondence of three successive nucleotides of mRNA (triplets) named codon to one amino acid residue (aa). Thus, 64 combinations of triplets correspond to 20 aa, which mean that different triplets can code for the same aa. This feature of the genetic code is known as degeneracy.

As a rule, triplets coding for the same aa differ from each other by the third base. To explain this, F. Crick suggested the wobble hypothesis [1]. He postulated that bases in the first and second codon positions form canonical Watson-Crick A-U and G-C, while in the third position non-canonical pairs can be formed, e.g., G-U pairs. Thus, one tRNA molecule could recognize two codons that differ by the last base. Some tRNAs contain in their anticodons inosine, which according to the Crick's wobble hypothesis is able to form pairs I-C, I-U and I-A. This suggestion has been made to explain triple degeneracy of isoleucine codons and also to explain why in some cases third base of triplet is not significant for the aa coding. The crick's hypothesis predicted specific geometry of the non-canonical base pairs G-U, I-U and I-A (I-C pair could be attributed to the canonical ones), therefore, this hypothesis can be examined with the use of structural data.

To date data have been accumulated that are not consistent with the Crick's wobble hypothesis. For example, it is known that bacterial isoleucine tRNAs do not contain inosine despite of the triple degeneracy of the Ile codons in bacteria. Structural basis of the wobble hypothesis was not confirmed too. Available X-ray data demonstrate that geometry of the G-U pair differs from that predicted by Crick. Fisrt, this has been shown with G-U pairs formed by modified uridine [2, 3], and then with the pairs within codon-anticodon duplex [4]. A number of known facts cannot be explained by the Crick's hypothesis, such as coding of Trp by the single codon, presence of modified uracyl in the first codon position, and, the most significant, the symmetry of the genetic code table reported by Rumer as early as in 1966 [5]. Double and four-fold degenerate codons are positioned in the table not randomly. Exactly in ½ of triples the third base is not coding, and it is insignificant whether it is purine or pyrimidine.

| II<br>I | C | G | U | A |
|---|---|---|---|---|
| C | Pro | Arg | Leu | Gln<br>Hys |
| G | Ala | Gly | Val | Glu<br>Asp |
| U | Ser | Trp<br>Cys | Leu<br>Phe | Stop<br>Tyr |
| A | Thr | Arg<br>Ser | Met<br>Ile | Lys<br>Asn |

Figure 1. Symmetry in the genetic code table according to Rumer [5]. "Strong" roots (marked as grey) are behind the diagonal. This form of the table is the most suitable to detect Lagerkvist's rules.

According to Rumer, the genetic code table can be written so that all these triplets are positioned above the diagonal (Fig. 1), providing a regularity, which is in general somewhat similar to Mendeleev's periodic table of elements.

Degeneracy in the genetic code was explained by U. Lagerkvist [6], who suggested that in each mRNA triplet only the first and the second bases are coding. This hypothesis did not directly contradict the wobble hypothesis, and did not make any structural predictions. Here, a new hypothesis is suggested that is in general consistence with the Lagerkvist's one, and can explain the structure of the genetic code and physic-chemical mechanism of the coding. The suggested molecular mechanism can explain the Rumer symmetry of the genetic code table, including coding of Met and Trp by the single codons, but abandons the Crick's structure of G-U pair.

**1. Chemistry and the wobble hypothesis**
*1.1. Wobble hypothesis and state-of-art experimental data*

The worldwide known Crick's wobble hypothesis [1] can explain formation of G-U pair when the third codon base is paired with an anticodon and coding of Ile by three codons (in the universal genetic code). The hypothesis is based on two suggestions: (i) nucleotides in the third codon positions can wobble, i.e., form base pair as a result of significant change of the canonical geometry, and (ii) inosine can form base pairs I-C, I-U and I-A (the structure of wobble G-U pair as well as the structures for I-U and A-I pairs were also suggested). However, both suggestions are not necessary to explain the known facts. Indeed, pair G-U has been actually detected [7], and the Crick's structure for this pair have been observed by X-ray [8], and purine-purine pairs (e.g., a pair 8-oxoguanine-adenine) were found to exist in the DNA double strand [9]. But these confirmations were obtained in model systems dissimilar to codon-anticodon duplexes, therefore could not be considered as confirmations of the wobble hypothesis.

The wobble hypothesis has been criticized several times in different reports since 1971 [6, 10, 11], and the most known is already mentioned work by U. Lagerkvist [6]. The latter was the first report showing that some features of the genetic code table could not be explained in the frames of Crick's wobble hypothesis. Various data available to date indicate that molecular mechanism of the G-U pair formation differs from that suggested by Crick [2, 3, 4, 12, 13], which in other words means partial reconsideration of molecular basis of the genetic coding.

*1.2. "Purine-purine wobble pair"*

Surprisingly but well known data contradicting the Crick's hypothesis did not lead to its reconsideration. As already mentioned, bacterial isoleucine tRNAs do not contain inosine [14] despite of three Ile codons in bacteria. Detailed study of bacterial isoleucine tRNA containing minor base lysidine complementary to adenine [15] showed that the pair -adenine is formed without wobbling, with very low probability of mispairing of lysidine with C or U. Besides, direct experiments showed that A-I pair destabilizes codon-anticodon interaction and is ineffective in coding [16]. The mentioned data imply that A-I pair is not necessary to explain triple degeneracy in the genetic code for Ile as well as anticodon wobble. Lysidine recognized A and does not misread U or C; nevertheless, the lysidine-containing tRNA is an isoleucine one together with Ile tRNAs containing G in the first anticodon position. In bacteria, inosine was found only in arginine tRNAs, therefore in frames in the Crick's hypothesis it is impossible to explain coding properties of triplets where only two bases are significant (e.g., all triplets GGN code for Gly, although inosine in glycine tRNAs was never found to the author's knowledge).

*1.3. Alternative variants of G-U pair structure*

The most known wobble Crick's structure of the G-U pair is tightly bound to the topic of the genetic code degeneracy. However, earlier F. Crick and J. Watson in their famous work hypothesized that tautomer forms of nucleotides could be responsible for misreading [17]. This implies possible existence of G-U pair where either the G or the U is in enol form. Therefore the later studies of tautomer forms of the bases were related to mutagenesis mechanisms [18-20]. An alternative variant is an ionized G-U⁻ pair, which was discussed in several works devoted to properties of modified nucleotides [21, 22]. Notably, both variants were considered for structures of G-U pair in codon-anticodon duplexes [3, 4, 12].

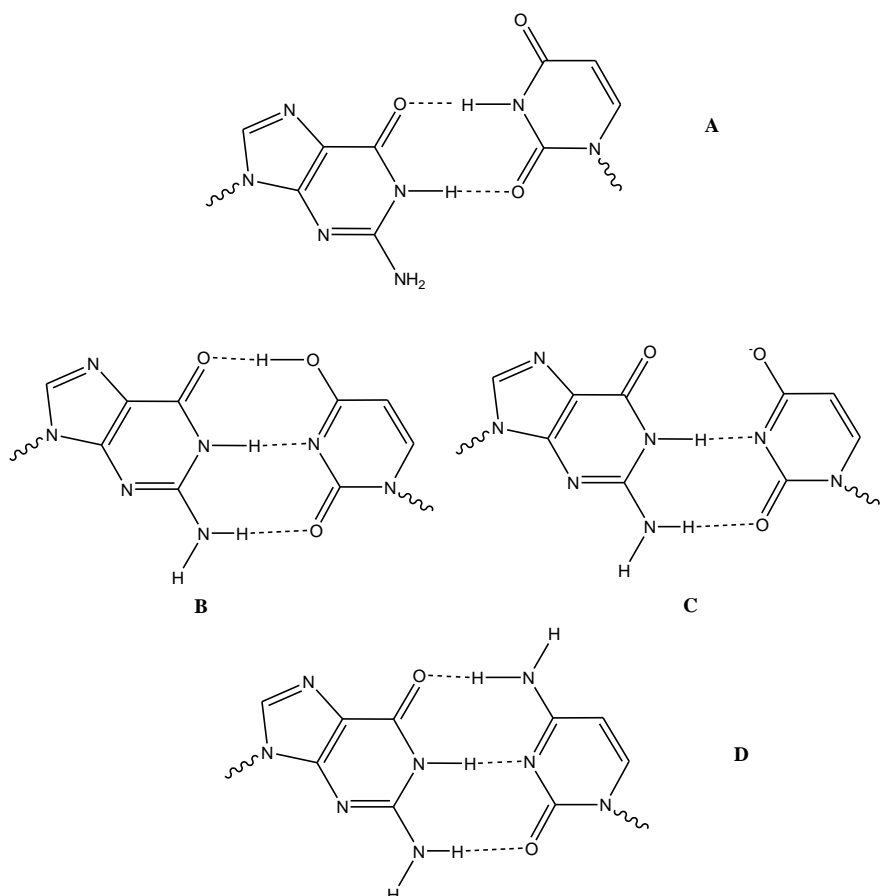

Figure 2. Three alternative structures of the G-U pair. (A) The Crick's wobble structure, (B) structure with enol form of the U and (C) structure with ionized form of the U. Structures B and C are similar to the canonical Watson-Crick C-C pair (D).

The alternative G-U pair structures are similar to each other (see Fig. 2) and have an advantage versus the Crick's wobble structure since they resemble canonical G-C pair and do not disrupt the double helix structure in contrast to the wobble structure (the latter has been originally proposed to disrupt the double helix structure). In this case, G-U pair becomes an analog of G-C pair providing a possibility of coding the same aa by different triplets varying in the third base. The Crick's wobble hypothesis ignores stacking between the neighboring bases. The stacking depends on the overlapping area of π-orbitals of the heterocycles. Therefore, disruption of the double helix structure leads to a loss because of less stacking with the Crick's wobble G-U structure. The loss could be compensated by the formation of an additional hydrogen bond, but the alternative structures also have two or even three hydrogen bonds. Taken all this into account, the alternative structures seem preferable as compared to the Crick's wobble G-U structure.

### 1.4. Modified uracil derivatives in G-U pairs

A base in the first anticodon position is often modified [23, 24]. In particular, a number of modified uracil derivatives have been found in this position. This gives some reasons to reconsider canonical ideas of base pairing. So, Björk et al. [13] showed that uridine-5-oxyacetic acid significantly changes character of codons recognition and suggested to revise the wobble hypothesis. Agris et al. demonstrated that modified base 5-methoxycarbonylmethyl-2-thiouridine in the first anticodon position is in a enol form in the pair with G [2, 3]. The authors suggested to widen the wobble rules taking into account features of modified bases [25-27]. Takai et al. [12, 28, 29] studied modified uracyl derivatives in the first anticodon position and obtained data indicating that pair of these derivatives with G differs from the Crick's wobble G-U structure and ionized form of the pairs was suggested. Altogether, in the works [25-27, 13, 30] the Crick's wobble G-U structure is not excluded from the consideration, but only suggest existence of other structures in the cases of modified nucleotides.

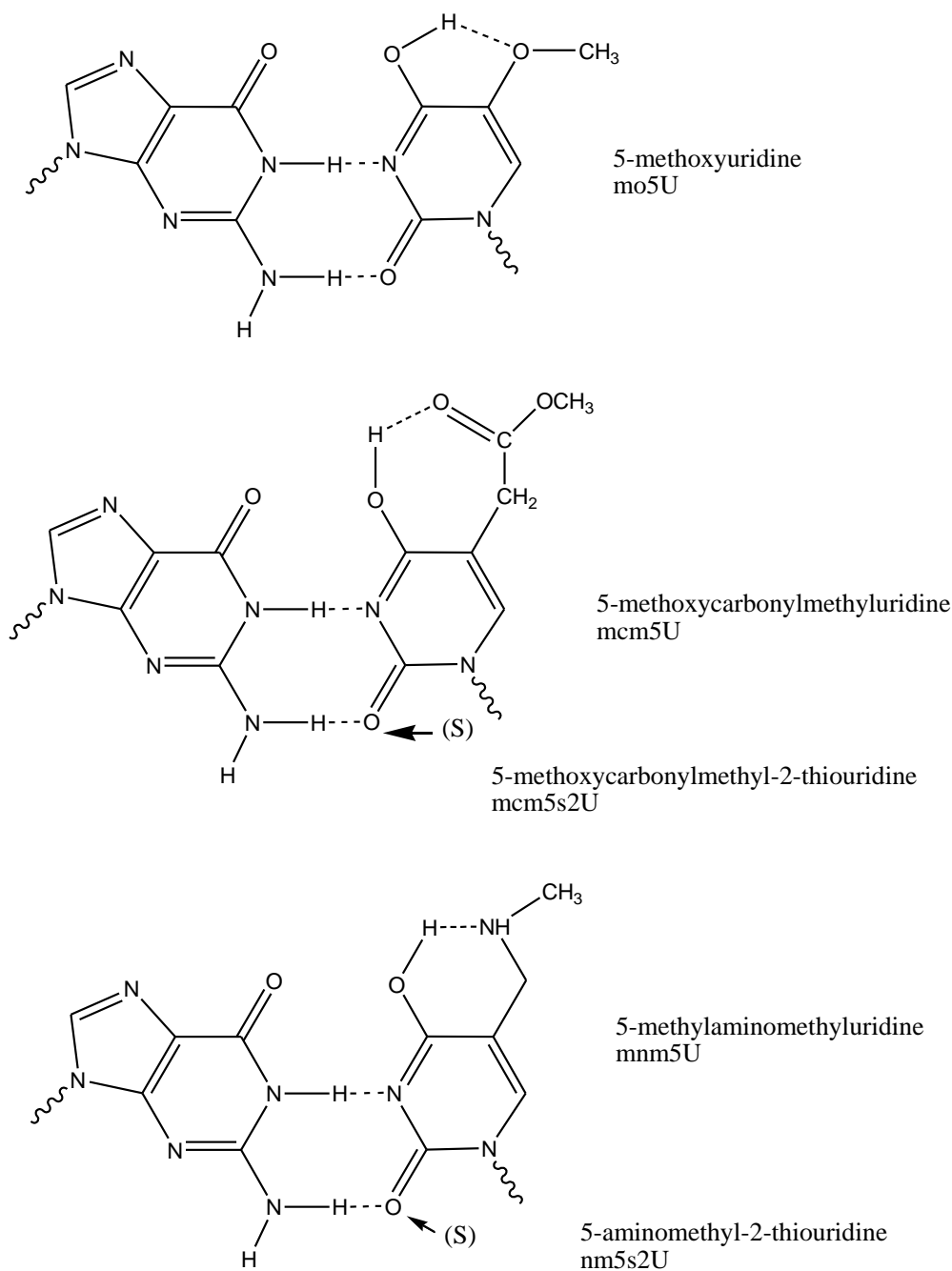

Figure 3. Three types of uracyl modifications often found in first position of anticodon. The structures demonstrate possible stabilization of the enol form of uraci;, which should in turn facilitate G-U pairs formation.

A number of modified bases that were found in first anticodon position can switch over the enol form easier than unmodified uracil. This possibility is illustrated by the structure of the uracil derivatives (Fig. 3) and is in a god agreement with biochemical data [2-3, 12-13, 24-28]. Stabilization of the enol form most probably occurs due to formation of an intramolecular hydrogen bond, which gives an additional support for the suggestion on non-wobble structure of G-U pairs formed by uridines in third position of codon. It is reasonable to suggest that all modifications of U in the first anticodon position facilitate formation of G-U pair structure similar to that of G-C pair.

The suggested mechanism for appearance of the enol form with modified uracil derivatives proceeds via intermediate states shown in Fig. 3. Formation of the third hydrogen bond in the GU-enol pair would disrupt intermolecular hydrogen bond in the modified uracil. If so, substituent in the uracil can somewhat decrease stability of the GU-enol pair. The mentioned mechanism should increase rate of formation of GU-enol basepairs. Formation of the intermolecular hydrogen bond can occur only upon transition to the enol form, thus this transition can take place before complementary interactions. Two hydrogen bonds (as seen from Fig. 3) are enough for initial recognition excluding formation of GU-wobble pair. The enol pair is formed after proton transfer, subsequent disruption of the intermolecular hydrogen bond and formation of new third O-H-O hydrogen bond between the guanine and the uracil. The suggested mechanism also clarifies which contribution modification of uracil makes to the recognition of adenine in wobble codon position. So, relative stability of intermolecular hydrogen bonds in modified uracil derivatives complicates AU pairs formation [12- 13, 23-31] It should be noted that these complications should be of kinetic nature since almost all known substituents in uracil position 5 are electron donors and therefore increase thermodynamic stability of the keto form.

Notably, thymine present in DNA is 5-methyluracil, a modified uracil. Methyl group as electron donor is expected to prevent enol formation. Replacement of U with T should lead to more unambiguous nucleotides recognition, and probably only T is present in DNA to make complementary interactions maximally unambiguous. This is consistent with comparative data on stability of G-T and G-U pairs.

1.5. Guanine in the anticodon

For clarity, in this section we'll write complementary pairs as follows: the first letter is mRNA nucleotide in wobble codon position, and the second letter is the complementary nucleotide in tRNA anticodon. Let us consider a UG pair (U in a codon, G in an anticodon). The structure of this pair has been solved in the codon-anticodon duplex in the ribosome, and it has been shown to be UG-wobble [4]. On the other hand, it was detected that formation of UG pair decreases rate of translation [32, 33]. Notably, in half of human codon-anticodon families, UG pairs are formed by unmodified guanines in wobble anticodon positions.

In frames of the suggested hypothesis one can explain data concerning UG pairs. With modified uracil derivatives we have suggested that the modifications accelerate transition to the enol form (see above). By analogy, modified guanines in wobble anticodon positions have the same function. Fig 4. shows modified nucleosides quenosine and archaeosine that substitute guanosine in many bacterial and archaeal tRNAs, respectively. Both nucleosides can form intermediate structures facilitating transition to UG-enol form in the same way as discussed above with modified uracil derivatives. The same function of overcoming the kinetic barrier is also inherent to inosine, the best known substituent of guanine in wobble anticodon position. In the frame of the classic wobble hypothesis, the role of inosine is to explain triple degeneracy of the genetic code of isoleucine. As was shown in section 1.2, this explanation is incorrect, and here another role of inosine in translation is suggested

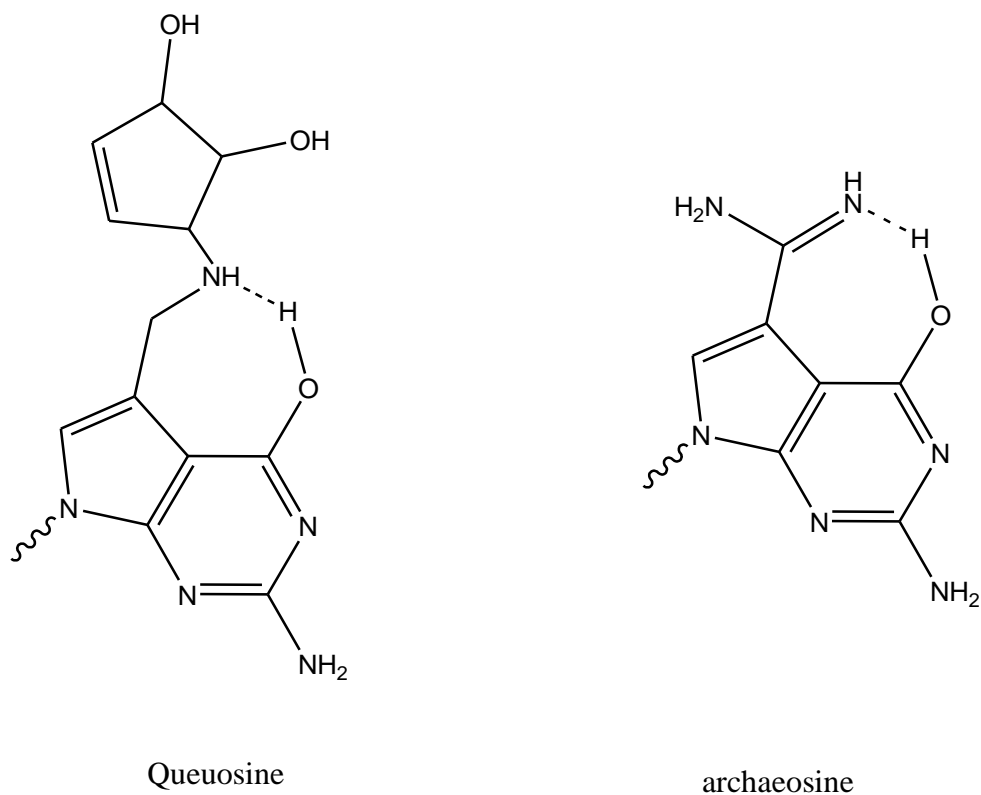

Queuosine    archaeosine

Figure 4. Guanine analogs in found tRNAs and intermediates facilitating their transition to enol form.

Let us consider inosine as guanine analog in the UG pair since inosine can be regarded as guanine without the amino group. One can assume that amino group complicates formation of UG-enol, e.g., this group can interact with ribosomal components and thereby stabilize UG pair in the wobble conformation, which has been observed in the X-ray structure of the 70S ribosomal complex. If so, absence of the amino group although should decrease thermodynamic stability of the UG-enol pair but simultaneously should facilitate transition of the UI pair to the enol form. It was demonstrated that between pairs CI and UI the difference in translation rate is less than that between pairs CG and UG [33], and translation rates with the mentioned basepairs are as CG>CI>UI>UG. In this case, as well as with uracil derivatives (see above), benefit is in the increased translation rate but not in stability of the suggested enol structure. It is seen that translation rate with CI pair is lower than that with CG pair, which resembles data with modified uracil derivatives in wobble anticodon position.

Thus, data on UG pairs are in agreement with the proposed hypothesis on the tautomeric structure of GU and UG pairs in wobble position.

*1.6. Enol or ionized form?*
All considerations on the effect of uracyl modifiecations on the stability of the G-U pairs discussed in the previous section relates equally to both enol and ionized structures. The suggested G-U pair geometry should also fit I-U pairs, but in the latter case only with enol form since ionized inosine can form only single hydrogen bond. On the other hand, the enol form was not detected in direct experiments with 5-halogeno-uracils and indications for the ionized form have been obtained [3, 34, 35]. Using similar methods and objects containing modified uracyl derivatives, both enol [2] and ionized [12] forms were reported. It is worth to mention here that X-ray crystallography, which is considered as a standard method for structure elucidation [4, 28] is unable to visualize protons, although this is of primary importance in the discussed problem.

G-U pairs are often present in RNA structures, e.g., in tRNAs. Ability of this pairs to bind a $Mg^{2+}$ ion [36, 37] can be considered as an indication for the ionized form. The same indication can be found in the data on rearrangements in RNA regions containing G-U pairs induced by pH alterations [7]. Charged G-U pair could contribute to the high selectivity of coding: one such pair in the "wobble" position should prevent formation of other ones involving the first and second nucleotides of the codon.

Keto-enol balance is generally thought to be pH-independent, while ionization evidently depends on pH. Thus, if a pH-dependence is found, involvement of ionized forms is expected, which is a simple test to distinguish enol and ionized forms. Such dependence has been found with model mutation, G-U pair in a DNA duplex bound to the active site of a DNA polymerase [38] pointing to the ionized form of the G-U pair. However, similar results with A-C pair were treated as indication for the involvement of a tautomeric base form in the pair formation [39]. However, the authors did not treat both discussed forms as mutually exclusive, and the ionized form can be observed as deprotonated enol one. Both forms can provide G-U pair geometry similar to that of G-C pair. The most informative approach to distinguish the enol and ionized forms could be NMR to locate specific protons. Tautomer and ionized forms of C and U derivatives were demonstrated in duplexes of oligodeoxyribonucleotides [21, 22, 34]. Unfortunately, these data could not be directly applied to RNA duplexes since RNA and DNA duplexes have different structure peculiarities. The hypothesis on the enol for of G-U pairs could be confirmed in NMR experiments with RNA duplexes.

NMR spectra of G-U pair in enol and wobble form should significantly differ from each other. The wobble form has two imino protons while the enol has only one (NHN pattern), the latter is analogous to that in AU and GC pairs. Special NMR approaches have been developed to identify NHN hydrogen bonding [40, 41] that are applicable both to Watson-Crick and wobble pairs. These approaches show that in two-dimensional 1H15N spectra signals of nitrogens correlate with each other at a close value of proton signal. So, close position of a proton to both nitrogens can be directly demonstrated, which is expected with G-U pair in enol form.

Correlation of two nitrogens belonging to the G and the U was observed in 1H15N NMR spectra of RNA duplexes and RNAs possessing a complex secondary structure [42-46]. In the mentioned studies the samples contained G-U pairs, and the correlation can be interpreted as a peculiar indication for the existence of the pairs in enol form. In spectra of tRNAs amount of such correlated signals even exceeds the amount of the G-U pairs [44-46]. These extra correlations could be assigned to G-Ψ pairs that always present in tRNAs. (See Supplemental material)

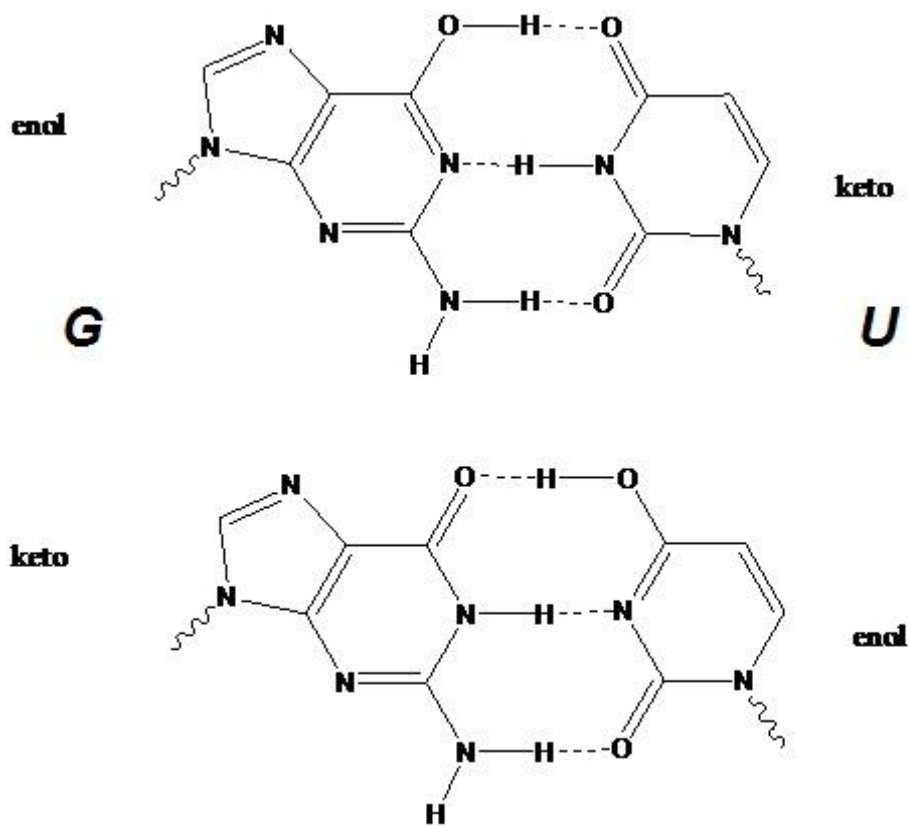

Figure 5. Two alternative variant of GU-enol structure.

The correlations of nitrogen signals of G and U in the enol form of G-U pair could be easily explained by the existence of two relatively stable structures (Fig. 5). The proton can move from one nitrogen to another, and this allows observation of both nitrogen signals as diagonal cross-peaks. This unique behavior could also be related to the specific environment of the G-U pair, in which uracil was in all cases in context 5'GUC3' or 5'CUC3'. If the discussed effect will be confirmed in experiments with G-U pairs selectively labeled with $^{15}$N in the duplexes, stable nature of enol form of G-U pairs will be demonstrated, and it will be a suitable approach to find such pairs in NMR spectra.

*1.7. Direct demonstration of non-wobble geometry of G-U pairs involving first and second codon nucleotides*

During translation tRNA can interact with a "near cognate" codon differing from cognate codon by one nucleotide [47] so that the first or the second codon nucleotide can be involved in the formation of a G-U pair. Recent X-ray analysis showed that geometry of these G-U pairs is not consistent with Crick's wobble structure but resembles that of the G-C pair [4]. These findings, together with data discussed above, allows reconsideration of the Crick's idea on the wobbling G-U pair in the third codon position, though the authors of [4] emphasize that they observed wobble G-U pair at the third codon position. One would think that conditions of all three codon position are very similar, but G-U pairs resembled the geometry of G-C pairs only in the fist and the second codon positions but not in the third one. This could be due to minor-groove interaction [48] which stabilizes G-C pairs in the first and the second positions by hydrogen bonding with A1492 and A1493 of the 16S rRNA, which should be also with enol G-U pairs having the same geometry, and therefore the enol form is stabilized by three more hydrogen bonds as compared with the wobble ones. This stabilization does not occur with the pair in third position, however, even without it a close analog of G-U pair demonstrates enol form [2, 3]. The existence of the wobble pair in the third position in the crystal structure [4] does not exclude a possibility of occurrence of enol G-U pair under physiological conditions.

Summarizing the discussion, one can state that only one of Crick's postulates mentioned in the section 1.1 remains valid, namely, G-U pairs really exist. But it seems unlikely that nucleotides in third position of codon form pairs by their significant displacement from canonical Watson-Crick positions. Crick's hypothesis do not provide molecular basis to explain the work of codons where only two bases are significant, and it remains unclear what is the difference of third position (where G-U pairs are allowed) from first and second positions.

## 2. Molecular basis of the degeneracy in the genetic code table

This section provides an explanation why majority of amino acids are coded by several codons and why methionine and tryptophane are coded by single codons together with rationale for the symmetry table of the universal genetic code.

*2.1. Lagerkvist's hypothesis*

Lagerkvist reported his hypothesis as early as in 1978 [6] and it has been formulated in three rules: (i) if two first pairs of the codon-anticodon duplex are maintained by 6 hydrogen bonds, than third codon base is insignificant for coding; (ii) if two first pairs are maintained by 4 hydrogen bonds, then different triples code for different aa dependent on the nature of third codon base (purine or pyrimidine) and (iii) if two first pairs are maintained by 5 hydrogen bonds, then rule (i) or (ii) is valid with pyrimidine or purine, respectively in the second codon position. These rules are beyond the frames of the Crick's hypothesis, which does not account to the number of hydrogen bonds formed by first and second nucleotides of codon and to the nature of the second codon nucleotide. In general, Lagerkvist suggested that two codon nucleotides are enough for coding, and his hypothesis became known as hypothesis "two out of three"; Lagerkvist's rules can be seen in the genetic code table suggested by Rumer [6] (Fig. 1). He developed his conception for a long time trying to popularize his main conclusion [49-53]. Nevertheless, Lagerkvist's hypothesis did not adequately replace the Crick's one because the latter made specific structural predictions for nucleotide pairs in contrast to the former one. Here, it is suggested molecular mechanism by which Lagerkvist's rules work. The essence of the suggested molecular mechanism can be formulated as a statement "Coding properties of a codon are completely defined by conformation of the first and the second bases, and this is the only thing recognized in the course of translation". In other words, both Lagerkvist's rules and Rumer's symmetry are explained via the structure of the codon-anticodon duplex.

*2.2. Rumer's symmetry of the genetic code table*

Lagerkvist when formulated his rules [6] did not referred to the earlier Rumer's work, but it is evident that both authors actually described the same phenomenon. The main thing in the Rumer's table (Fig. 1) is the existence of "roots", i.e. two first nucleotides of codon and their ability/inability to code for only one aa [5, 54, 55]. Of 16 roots, 8 are "strong" (coding for single aa) and 8 are "weak" (coding for more than one aa). Rumer hoped that the discovered regularity will be explained in frames of molecular mechanism of decoding; the same hope was repeated in later work [56] where the genetic code table symmetry was described again. The reported symmetry is typical with all genetic code dialects, and this symmetry could not be deduced from rules defined by the Crick's hypothesis. The symmetry of the genetic code table was a subject of several theoretical papers [57-62], but molecular mechanisms of coding and molecular basis of the symmetry were out of consideration in these papers. Surprisingly, theoretic works on symmetry of the genetic code developed separately from the studies on molecular mechanisms of decoding. So, an attempt to explain Lagerkvist's rules via interactions of codon-anticodon duplex with the ribosomal RNA has been made [63]. In particular, it was suggested that involvement of two first codon bases in so called A minor interactions is responsible for their coding properties. However, the relationship between the A minor interactions and coding properties of the first codon bases were not in fact argued, and the symmetry of the genetic code table was not mentioned.

Here in the next section a universal mechanism of recognition of all codons is suggested. This mechanism does not require redundant assumptions and could be the actual molecular basis for the Rumer's symmetry and Lagerkvist's rules.

*2.3. Conformation of codon*

In the above section evidence was presented that the structure of the G-U pair is close to the canonical Watson-Crick pairs, therefore, the main attention is paid to possible conformational rearrangements in codon-anticodon duplexes that can occur without hydrogen bonds disruption. Rumer's symmetry can be explained if one assumes that mutual arrangement of the first and the second codon bases is dependent on the nature of the third one (keeping in mind the arrangement of codon-anticodon duplex but not a trinucleotide in solution or triplet in free mRNA).

Conformation of the duplex is defined by hydrogen bonding of complementary bases and stacking of neighboring bases [64]. Stacking is an unspecific phenomenon [65] and its strength decreases in the range (purine-purine)>(purine-pyrimidine)>(pyrimidine-pyrimidine). The less distance between the interacting bases, the higher stacking; besides, the strength of the stacking depends on the angle between the interacting bases since it is defined by overlapping of the π-electron systems. Thus, stacking can both bring bases close to each other and rotate them. It is easy to observe that the number of hydrogen bonds formed by the third codon base is almost insignificant, while its nature (purine or pyrimidine) is essential. Thus, it is reasonable to relate the existence of "strong" and "weak" roots in the Rumer's table (Fig. 1) to the stacking. Double helix structure has limited degree of freedom and it is difficult to imagine that hydrogen bonds affect one part of the molecule and the stacking on another. Thus, the increased stacking between the second and the third codon bases can affect the arrangement of the second base changing its positioning regarding the first base, which can be referred further as "conformational changes of the codon root". The main assumption of the scheme suggested here is that the "strength" of the codon root is exactly and specifically concerned with its conformation, and that the conformation of strong roots does not depend on the nature of the third codon base.

In the codons with roots CC, CG, GC and GG the root conformation is completely defined by the complementary interactions. Three hydrogen bonds in the each pair make conformational alterations barely possible. These cases are shown in the respective corner (shaded by deep grey) in Fig 6.

The next step is analysis of the root pairs (UC-UG), (AC-AG), (CU-CA) and (GU-GA) (shaded by light grey in Fig 6.). In each pair the first root in strong and the second is weak (i.e., the coding properties of a codon UCN do not depend on the nature of N, while those of UGN depend). In each pair the number of hydrogen bonds is the same, but the weak root is that which has purine in the second position. The latter allows occurrence two successive purines in the second and the third positions, which provides the maximum stacking that in turn alter mutual positioning of the first and the second bases. It should be noted here that anticodon is less conformationally flexible than codon since it is located within rather rigid tRNA structure. The same arguments are applicable to the rest root pairs (AC-AG), (CU-CA) and (GU-GA).

The analysis of the table is completed with the consideration of its last quarter containing the roots whose nucleotides form only four hydrogen bonds. In codons with the roots UA and AA the presence of a purine in the third position allows alteration of the root conformation, and in codons with roots UU and AU weaker pyrimidine-purine interaction is sufficient to change the root conformation. Here it is implied that stacking Py-Pu stronger affects the root conformation than stacking Pu-Py, because in codons with the roots UA and AA the stacking Pu-Py does not result in the change of the root conformation. The mechanistic illustration of the suggested model is presented in Fig.7 where on one scale is the number of hydrogen bonds fored by the codon root and on another the strength of stacking between the second and the third bases.

|   | C | G | U | A |
|---|---|---|---|---|
| C | <u>III III</u> Pu<br>          }Pro<br><u>III III</u> Py | <u>III III</u> Pu<br>          }Arg<br><u>III III</u> Py | <u>III II</u> Pu<br>          }Leu<br><u>III II</u> Py | <u>III</u> **II Pu**   Gln<br><br><u>III II</u> Py   Hys |
| G | <u>III III</u> Pu<br>          }Ala<br><u>III III</u> Py | <u>III III</u> Pu<br>          }Gly<br><u>III III</u> Py | <u>III II</u> Pu<br>          }Val<br><u>III II</u> Py | <u>III</u> **II Pu**   Glu<br><br><u>III II</u> Py   Asp |
| U | <u>II III</u> Pu<br>          }Ser<br><u>II III</u> Py | <u>II</u> **III Pu**   Trp<br><br><u>II III</u> Py   Cys | <u>II</u> **II Pu**   Leu<br><br><u>II II</u> Py   Phe | <u>II</u> **II Pu**   Stop<br><br><u>II II</u> Py   Tyr |
| A | <u>II III</u> Pu<br>          }Thr<br><u>II III</u> Py | <u>II</u> **III Pu**   Arg<br><br><u>II III</u> Py   Ser | <u>II</u> **II Pu**   Met<br><br><u>II II</u> Py   Ile | <u>II</u> **II Pu**   Lys<br><br><u>II II</u> Py   Asn |

Figure 6. Degeneracy in the genetic code table is defined by conformational alterations of the first and second codon bases. Roman numerals indicate the number of hydrogen bonds formed by these bases, bold shows cases where stacking between the second and the third codon base can affect conformation of the first and the second bases. Pu, purine; Py, pyrimidine. The table is dissected into 4 blocks (marked with different shades of grey), which is suitable to consider molecular mechanisms underlying degeneracy in the genetic code.

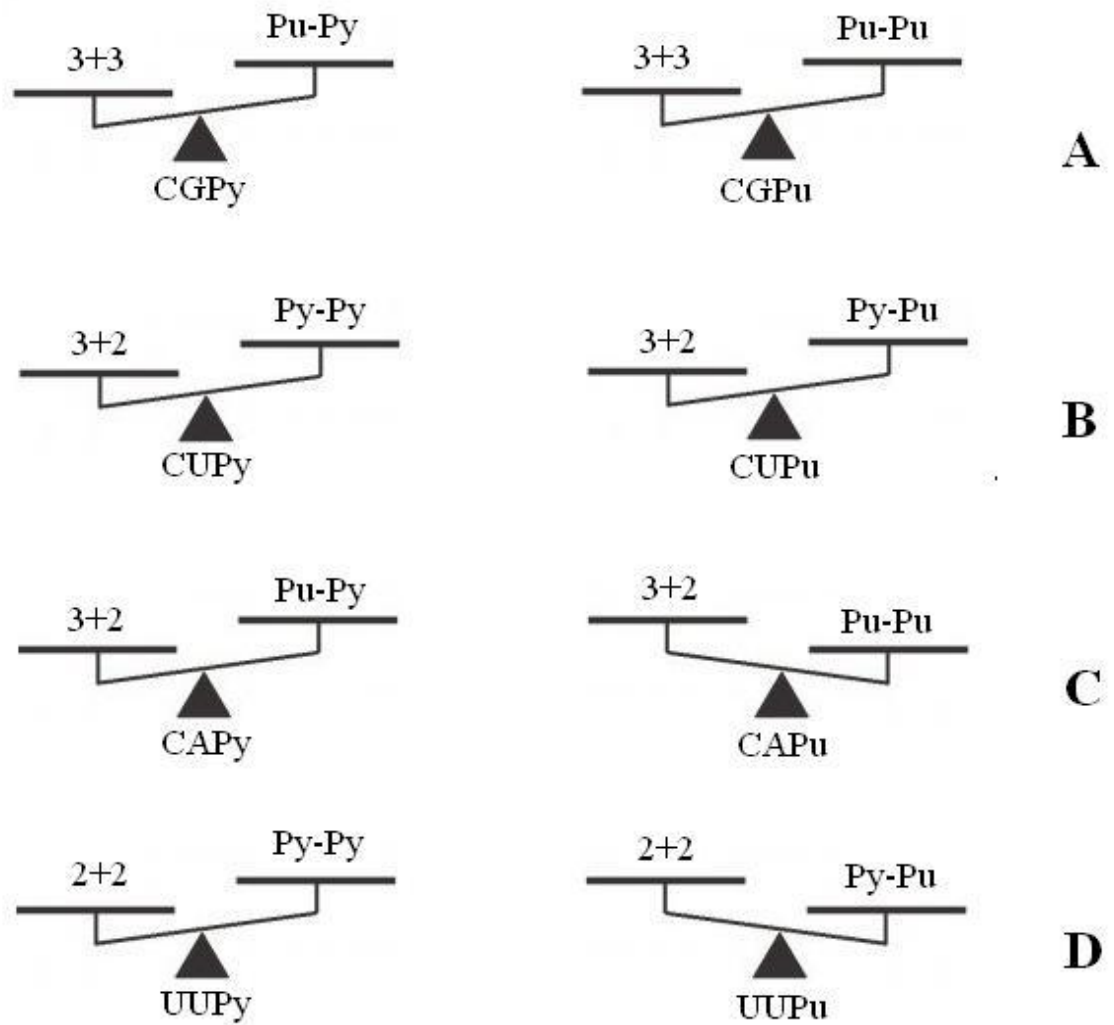

Figure 7. The effect of stacking between the second and the third codon bases on the codon root conformation. On the left scale is the number of hydrogen bonds formed by the root, and on the right stacking between the second and the third codon bases. **(A)** Stacking cannot change conformation of a root that forms 6 hydrogen bonds. **(B)** Stacking Py-Pu cannot change conformation of a root that forms 5 hydrogen bonds, but stacking Pu-Pu **(C)** changes the root conformation. **(D)** Conformation of a root that form 4 hydrogen bonds can be changed by stacking Py-Pu.

The suggested mechanism of the genetic code functioning allows exact identification of a thing that is common to codons varying in one nucleotide in the third position and coding for the same aa. This common thing is mutual spatial positioning of the first and the second codon bases, and this is an actual coding unit of the codon. With ½ of the roots this positioning depends on the third codon base, and this positioning is recognized but not the triplet sequence. This mechanism explains Rumer's symmetry of the genetic table and can be experimentally examined.

*2.4. Difference between the third codon position and the first two positions*

As mentioned above, G-U pair can be formed only by third codon base. Although this pair is structurally similar to the canonical G-C pair, it is recognized in other way than the G-C pair when occurs in the first or the second codon position. This does not contradict to the above suggestion that the recognized unit is conformation of the codon root. Although G-U pair is structurally similar to the G-C one, it is less stable, like A-U pair, but A-U pair has geometry dissimilar to the G-U pair. Thus, G-U pair in the first/second codon position is distinguished from a G-C pair by its stability and from a A-U pair by its geometry. The different stabilities should be recognized at the kinetic proofreading step, while at the step of initial recognition G-U and G-C pairs are not distinguished [4].

The main thing that makes third codon position different from the first two positions is that stacking of a base in this position with the neighboring base in the second position can affect conformation of the codon root and thus affect coding specificity of the triplet.

*2.5. Structural basis of the hypothesis*

Surprisingly, conformational alterations analogous to those described above have been reported earlier. RNA double helix can exist in two forms, A and A'. These forms differ by the number of nucleotides par one turn (11 and 12, respectively), by the pitch (30 and 36 angstroms) and by the angles between nucleotides; other parameters of these forms are the same [64]. Notably, contribution of stacking in the double helix formation in forms A and A' is significantly different. Possibly, differences in mutual positioning of the two nucleotide pairs formed by the codon root caused by replacement Pu to Py and vice versa in the third position are related to the mentioned alternative conformations of mini-helix. Progress in X-ray crystallography can provide a support for this assumption.

*2.6. Triptophane and methionine*

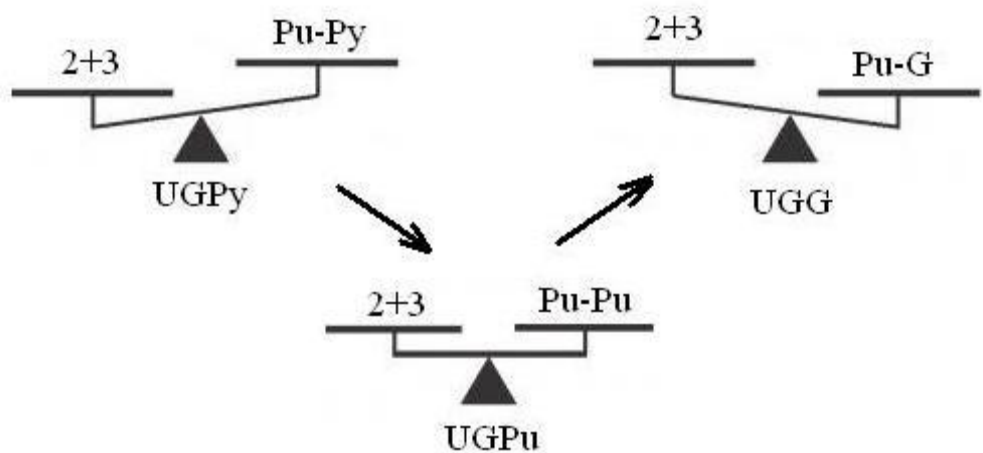

Figure 8. Root UG is close to an unstable equilibrium. Adenine in third position is unable to change the root conformation while guanine is able to do this. This explains the absence of degeneracy of the genetic code for Trp. Similar scheme concerns triplet AUG coding for Met.

In the universal genetic code two aa are encoded by single codon, namely Trp (UGG codon) and Met (AUG codon). This phenomenon can be explained using mechanistic scheme presented above in Fig.8. As already mentioned, conformation of some codon roots is not dependent on the nature of third codon base, while conformation of other root are dependent on it. In these terms, roots can exist that are close to an "equilibrium point". If balances are at equilibrium, a small addition (a minor effect that is not taken into account in other cases) to each scale changes the equilibrium. The root UG lies at the table diagonal (Fig. 8) and therefore could be regarded as being located close to equilibrium. So, its conformation can be dependent on such detail as the nature of purine in the third position, and replacement of G to A in this position results in the change of the root conformation. Similar reasoning concerns the AU root.

Unambiguous recognition of UGG and AUG triplets as Trp and Met codons is supported by the presence of C in the first anticodon position of the respective tRNAs in all organisms with universal genetic code. Evidently, a loss of a single hydrogen bond makes these anticodons invalid, and interactions in the scale of one hydrogen bond turns out to be of principal importance for the proper recognition of the mentioned two triplets during translation.

## 3. Conclusion

Molecular mechanism of the genetic code reading is suggested. This mechanism is based on four statements:
- pair G-U can be formed by third codon base, which lead to ambiguity of translation (this had been originally stated by F. Crick');
- the geometry of pair G-U is similar to that of G-C, which contradicts Crick's wobble hypothesis but is in accordance with Watson & Crick's note on a possibility of G-U pair formation via tautomer forms of the bases;
- the genetic code table is structured in such way that all codons in which third nucleotide is not coding are above the diagonal (such presentation had been originally reported by Rumer);
- degeneracy of codons depends on the number of hydrogen bonds formed by bases in the first and the second codon positions so that the nature of the coded amino acid is defined only by the first and the second codon bases (this had been originally proposed by Lagerkvist).

According to the mechanism proposed here, the unit to be recognized in the codon is mutual spatial positioning of the first and the second bases. This explains the difference between these codon position from the third one. The contribution of the third base into coding is via stacking with the base in the second position. The mechanism explains degeneracy in the genetic code, Rumer's symmetry of the genetic code table and Lagerkvist's rules as well as coding of methionine and tryptophan by single codons. The presented ideas on the structure of G-U pair allow understanding the function of uridine modifications in the first anticodon position in tRNAs as well as significant difference between thymine and uracil in their abilities to for base pair with guanine.

**Acknowledgements**

The author is grateful to D.M. Graifer for helpful discussion and his help in the text preparation.

Supplemental material.

**Five examples of G-U-enol (G-Ψ-enol) pair in NMR spectra**

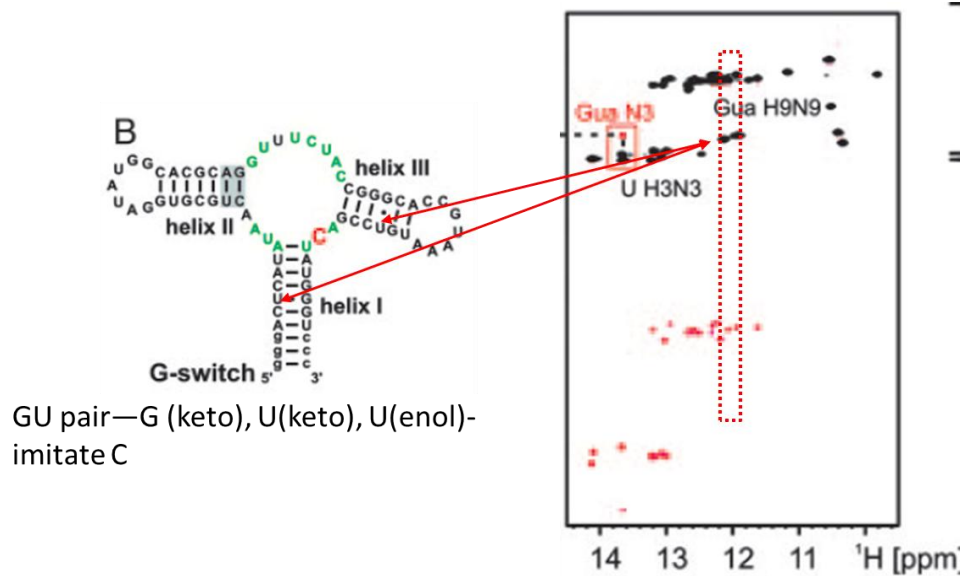

Suppl. Figure 1. J. Noeske at al. An intermolecular base triple as the basis of ligand specificity and affinity in the guanine- and adenine-sensing riboswitch RNAs. PNAS v. 102  5 1372–1377 (2005)

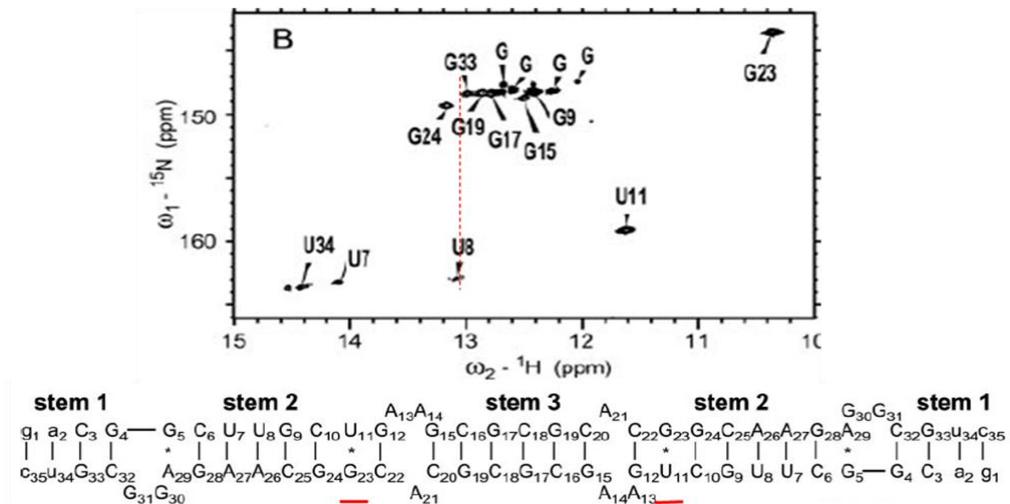

Suppl. Figure 2. N.B. Ulyanov at al. NMR Structure of the Full-length Linear Dimer of Stem-Loop-1 RNA in the HIV-1 Dimer Initiation Site. JMB 281 23 16168–16177 (2006)

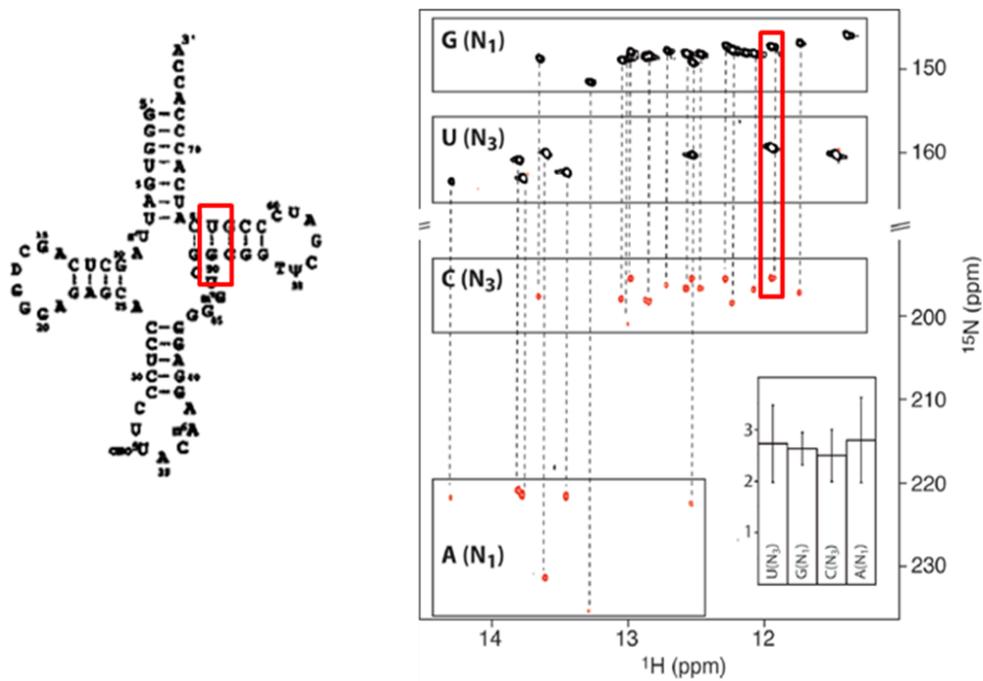

Suppl. Figure 3. J. Farjon at al. Longitudinal-Relaxation-Enhanced NMR Experiments for the Study of Nucleic Acids in Solution. J. AM. CHEM. SOC. 2009, 131, 8571–8577

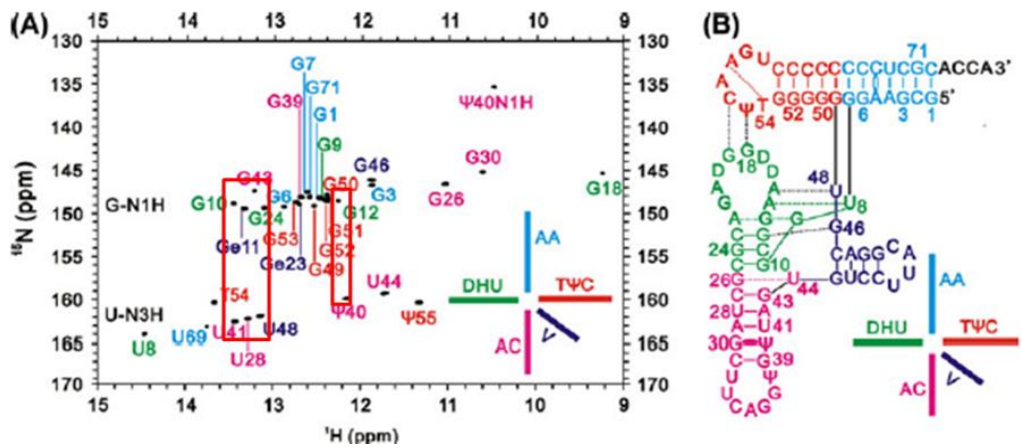

Suppl. Figure 4. Zhan-Xi Hao at al. 1H, 15N chemical shift assignments of the imino groups in the base pairs of Escherichia coli tRNALeu (CAG) Biomol NMR Assign (2011) 5:71–74

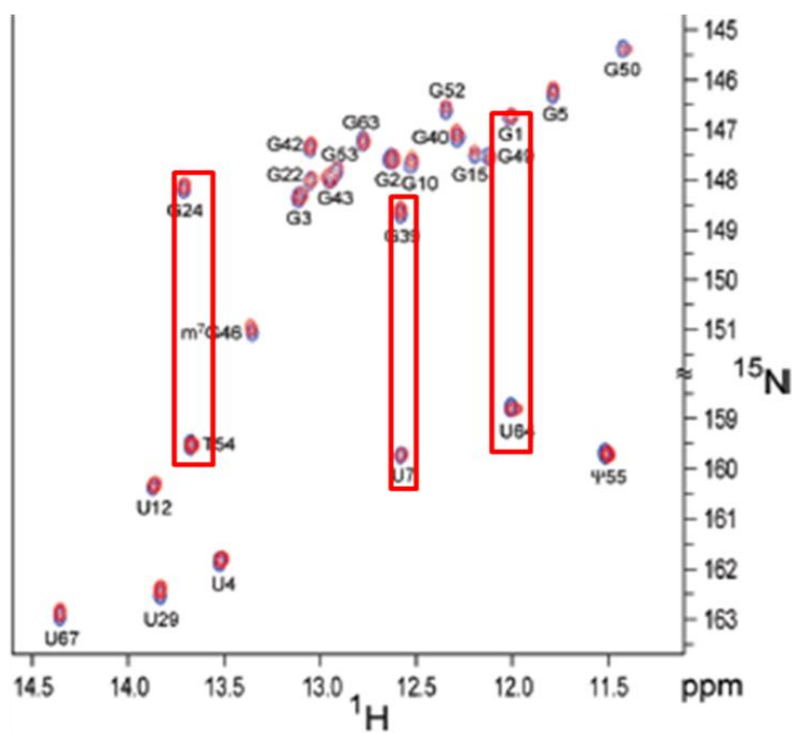

Suppl. Figure 5. Grishaev at al. Chemical Shift Anisotropy of Imino 15N Nuclei in Watson-Crick Base Pairs from Magic Angle Spinning Liquid Crystal NMR and Nuclear Spin Relaxation. J. AM. CHEM. SOC. 2009, 131, 9490–9491